%% file: main.tex
\documentclass[%
 reprint,
superscriptaddress,
 amsmath,amssymb,
 aps,
prd,
]{revtex4-2}

\usepackage{graphicx}
\usepackage{dcolumn}
\usepackage{bm}
\usepackage[colorlinks=true,citecolor=blue,filecolor=blue,linkcolor=blue,urlcolor=blue,luatex]{hyperref}
\usepackage[dvipsnames]{xcolor}
\usepackage[capitalize] {cleveref}
\usepackage{orcidlink}
\usepackage[inline]{enumitem}

\input{mydefs}
\begin{document}


\title{Mechanical Sensors for Ultraheavy Dark Matter Searches via Long-range Forces}

\include{authors}

\date{\today}

\begin{abstract}
Dark matter candidates with masses around the Planck-scale are theoretically well-motivated,
and it has been suggested that it might be possible to search for dark matter solely via gravitational interactions in this mass range. In this work, we explore the pathway towards searching for dark matter candidates with masses around the Planck-scale using mechanical sensors while considering realistic experimental constraints, and develop analysis techniques needed to conduct such searches. These dark matter particles are expected to leave tracks as their signature in mechanical sensor arrays, and we show that we can effectively search for such tracks using statistical approaches to track-finding.
We analyze a range of possible experimental setups and compute sensitivity projections for searches for ultraheavy dark matter coupling to the Standard Model via long-range forces.
We find that while a search for Planck-scale dark matter purely via gravitational couplings would be exceedingly difficult, requiring $\sim80 \unit{dB}$ of quantum noise reduction with a $100^3$ array of devices, there is a wide range of currently unexplored dark matter candidates which can be searched for with already existing or near-term experimental platforms.
\end{abstract}

\maketitle

\section{Introduction}

There is a large body of evidence supporting the existence of dark matter, but its fundamental nature remains largely unknown~\cite{Workman:2022ynf}. Dark matter candidates remain diverse, with an approximate allowed mass range spanning $10^{-22}\unit{eV/c^2}$ to $5\, M_{\odot}$~\cite{Workman:2022ynf}, where $M_{\odot}\approx2\times10^{30}\unit{kg}$. To search for dark matter across such a large parameter space, various experimental approaches have been applied across the direct detection community. These include liquid noble element detectors~\cite{XENON:2023cxc, XENON:2023iku, LZ:2024zvo, LZ:2024psa, DEAP-3600:2017uua}, bubble chambers~\cite{Behnke:2016lsk, PICO:2023uff}, and magnetically levitated sensors~\cite{Amaral:2024rbj}, among others. One region of parameter space that has been the focus of many dedicated searches is the ultraheavy dark matter region, constituting particles of masses around or below the Planck mass, $m_\mathrm{Pl}=\sqrt{\hbar c/G} \sim 10^{19}\unit{GeV} \sim 20\unit{\mu g}$~\cite{XENON:2023iku,LZ:2024psa,Cappiello:2020lbk, DEAPCollaboration:2021raj}. 

For Planck-scale dark matter, we expect a particle flux of
\begin{equation}\label{eq:flux}
    \Phi_\chi = \frac{\rho_\chi}{m_\chi} v_0 \simeq 0.2/\unit{yr\,/ m^{2}}\, \left(\frac{m_{\mathrm{Pl}}}{m_\chi}\right)\,,
\end{equation}
where we have used $\rho_\chi \approx 0.3\,\mathrm{GeV\,cm^{-3}}$ and $v_0 \approx 238\,\mathrm{km\,s^{-1}}$~\cite{Baxter:2021pqo}. This implies that the Planck mass is an approximate upper bound for the masses that can be probed by metre-scale experiments~\cite{Carney:2022gse}. 
The region of mass parameter space around and below the Planck mass is also theoretically well-motivated~\cite{Blanco:2021yiy}. This region includes Planck-mass relics from black hole evaporation~\cite{Lehmann:2019zgt, Carr:2021bzv, Profumo:2024fxq}, WIMPZillas~\cite{Kolb:1998ki, Kolb:2007vd, Kolb:2017jvz, Harigaya:2016vda}, asymmetric dark matter nuggets~\cite{Gresham:2017cvl, Wise:2014jva}, and Q-balls~\cite{Enqvist:2001jd, Cotner:2016dhw}, among others~\cite{Bai:2018dxf, Blanco:2021yiy, Carney:2019pza, Kolb:2023ydq}. Many of these theories also involve additional forces; as such, searches with sensitivities that cannot reach the gravitational coupling strength are still valuable tests of theories of dark matter. 

Mechanical sensor arrays are a good candidate when searching for Planck-scale dark matter via non-contact interactions because it is feasible to build metre-scale arrays of sensors to capture necessary statistics despite the low flux, and because the detection threshold for a mechanical sensor applies to the momentum transferred to a bulk test mass, unlike in scintillation detectors where the threshold applies to microscopic volumes within the detector. This greater sensitivity to non-contact forces makes such mechanical sensor arrays complementary to liquid noble element detectors~\cite{XENON:2023cxc, XENON:2023iku, LZ:2024zvo, LZ:2024psa, DEAP-3600:2017uua}. In addition, mechanical sensor arrays can be versatile experiments for searches for physics beyond the Standard Model, including TeV-scale composite dark matter that couples to Standard Model particles via long-range interactions~\cite{Monteiro:2020wcb}, the relic neutrino background via weak torques~\cite{Arvanitaki:2022oby, Smith1983CoherentIO}, and ultralight dark matter~\cite{Amaral:2024rbj, Carney:2020xol}.

In this work, we discuss key considerations regarding the use of mechanical impulse sensors to search for Planck-scale dark matter with long-range couplings and project experimental sensitivities. We do this from an experimentalist perspective, including experimental considerations such as the look-elsewhere effect, and using Monte-Carlo simulations to evaluate the sensitivity, thus expanding on prior work~\cite{Carney:2019pza}. We find that searching for dark matter purely via its gravitational coupling is exceedingly difficult, requiring even greater quantum noise reduction than suggested in~\cite{Carney:2019pza}, and appears to be infeasible with any realistic detector in the foreseeable future. However, we also find that even current detectors in small arrays would be sensitive to a wide range of new dark matter candidates, and future versions could further expand the parameter space being probed.

We start with a discussion of the detector concept and data-analysis challenges in~\cref{sec:detector_concept}, followed by a brief overview of the hardware approaches to realising such sensor arrays in~\cref{sec:experimental_realisation}. In~\cref{sec:sensitivity}, we show sensitivity projections of various experimental realisations. In~\cref{sec:grav}, we discuss the possibility of probing dark matter via the gravitational interaction. Finally, we summarise our key findings and discuss the future outlook of this research direction in~\cref{sec:conclusion}.

\section{Detector Concept}\label{sec:detector_concept}
\begin{figure}[htbp]
    \centering
    \includegraphics[trim={4cm 5.8cm 4cm 3.7cm},clip,width=\columnwidth]{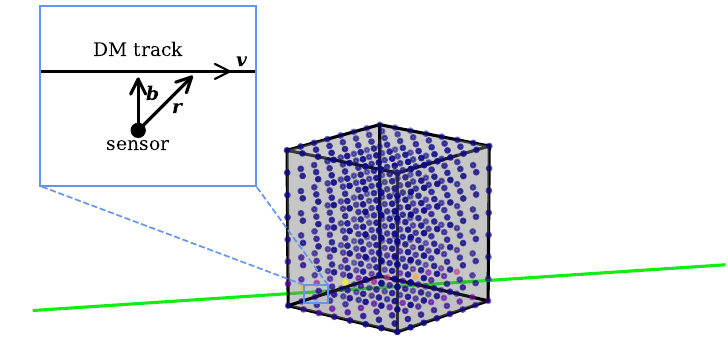}%
    \caption{Schematic of a 3D sensor array, where sensors are represented by coloured circles, with the colour indicating the strength of an acceleration signal. A dark matter track going through the array is shown in green. A zoomed-in schematic showing the impact parameter vector $\boldsymbol{b}$, the position vector $\boldsymbol{r}(t)$, and the dark matter velocity vector $\boldsymbol{v}$ is displayed on the top left.}\label{fig:windchime_array_plot_color}
\end{figure}

The detector concept is to have an array of mechanical impulse sensors. Assuming that dark matter interacts with the sensor array via a long-range force, a dark matter particle passing through the array would leave its signature as impulses along a track. Additionally, searching only for track-like signals with a velocity corresponding to what we expect from the local dark matter distribution would enable us to reject many backgrounds due to particles travelling through the array with different velocities versus dark matter~\cite{Carney:2019pza, Windchime:2022whs}; a similar strategy is proposed in~\cite{Gao:2025ryi} for the detection of millicharged dark matter. This method for detecting particles that interact with long-range forces should be similarly useful for detecting other particles that can couple via long-range forces, especially if the particles are non-relativistic. A schematic of such a sensor array is shown in~\cref{fig:windchime_array_plot_color}.

In this work, we consider ultraheavy dark matter interacting with baryonic matter via long-range forces. The interaction strength, $\alpha$, is parameterized via:
\begin{equation}\label{eq:force}
    F = \frac{\alpha \hbar c N_{\mathrm{nuclei}}}{r^2},
\end{equation}
where $N_{\mathrm{nuclei}}$ is the number of nuclei in the test mass and $r$ is the distance between the dark matter particle and the test mass. We parameterize the interaction strength based on nuclear number instead of nucleon number, $A$, because the typically assumed $A^4$ scaling for dark matter interaction cross section is not always valid for ultraheavy dark matter. Specifically, the assumptions required for $A^4$ scaling do not hold for large cross sections and non-contact interactions, and become model dependent~\cite{Digman:2019wdm}. While we study the detection of dark matter that couples to our sensor array via non-contact interactions, we also note that attractive forces interacting via a light mediator of mass $m_\phi$ would be indistinguishable from a long-range force as long as $1/m_\phi \gg d$, where $d$ is the sensor spacing.

For a dark matter particle passing by a single sensor with impact parameter $b$ and velocity $v$, the force is given by
\begin{equation}
    {F}(t) = \frac{\alpha \hbar c N_{\mathrm{nuclei}}}{(b^2 + v^2 t^2)^{3/2}}\left(\boldsymbol{b} + \boldsymbol{v}t\right)\cdot\hat{n},
\end{equation}
where $\hat{n}$ is a unit vector corresponding to the sensor's direction of sensitivity, $\boldsymbol{b}$ is a vector from the sensor to the point of closest approach of a dark matter track, $t$ is the time relative to the time of closest approach, and $\boldsymbol{v}$ is the velocity vector of the dark matter particle. We assume that the transferred momentum is small compared to the total momentum of the dark matter particle, such that $\boldsymbol{v}$ remains constant as the particle passes through our sensor array.

For simplicity, we consider only the force in the $\boldsymbol{b}$ direction; this is conservative as we neglect the force in the direction of the track, which is harder to detect due to there being no net impulse when integrated over time. If one considers a time-integral centred around the time of closest approach, the momentum transfer is
\begin{equation}\label{eq:momentum_transfer}
\begin{split}
    \Delta p &= \int^{b/v}_{-b/v}  \frac{\alpha \hbar c N_{\mathrm{nuclei}}}{(b^2 + v^2 t^2)^{3/2}} b (\hat{b} \cdot \hat{n}) dt\\ 
    &= \frac{\sqrt{2} \alpha \hbar c N_{\mathrm{nuclei}}}{b v}(\hat{b} \cdot \hat{n}).
\end{split}
\end{equation}
This impulse is the key observable in our sensor array. However, to improve sensitivity and improve background rejection, we aim to perform track-finding in sensor arrays, allowing us to observe tracks that might not produce statistically significant pulses in any individual sensor~\cite{Carney:2019pza, Qin:2023urf}. This is expanded upon in~\cref{sec:data-analysis}.

\subsection{Data Analysis}\label{sec:data-analysis}

\subsubsection{Statistical Track-finding}\label{sec:track-finding}

\begin{figure}[htbp]
    \centering
    \includegraphics[width=\columnwidth]{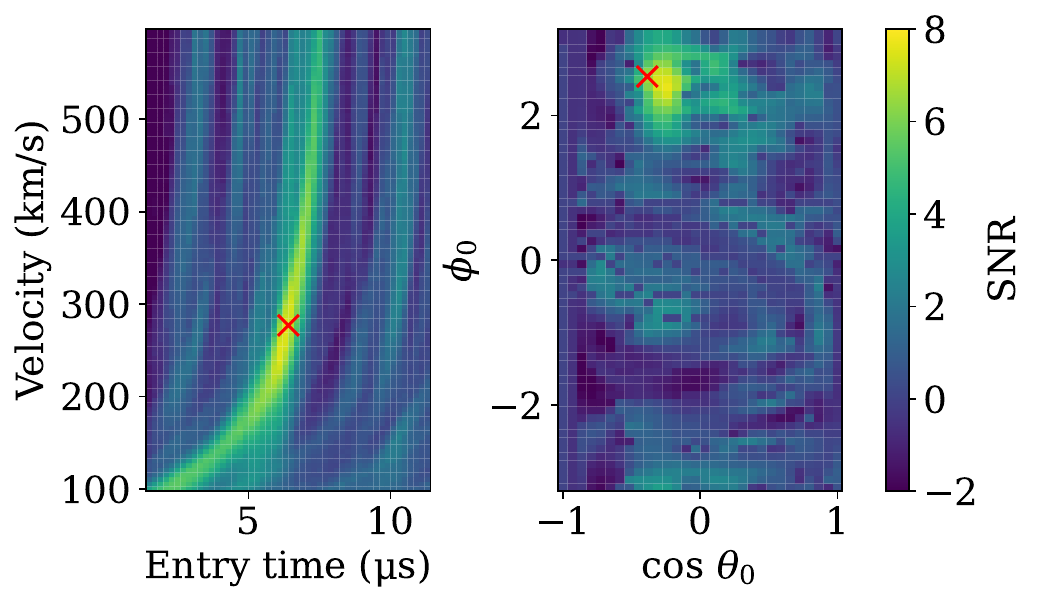}%
    \caption{2D slices of the template matching significance map. The signal-to-noise ratio (SNR) is shown by the colormap, and the truth values are shown by the red crosses. $\phi_0$ and $\cos \theta_0$ refer to the spherical coordinates of the entry point of the dark matter track. It can be seen that a clear peak around the truth parameter values of the track is clearly visible. These 2D slices are produced by setting all other parameters to the simulation truth values. }\label{fig:2D_SNR}
\end{figure}

Track-finding is a commonly-performed data analysis and reconstruction task in particle physics experiments~\cite{ATLAS:2021tfo, CMS:2014pgm}. Typically, tracks are detected and reconstructed in a sensor array based on statistically significant signals on individual sensors, often termed ``hits". In our proposed experiments, however, we want to take advantage of the signal summed across multiple sensors in the array to boost our sensitivity. A track that is only statistically significant after summing up the signals from multiple sensors cannot be detected by performing track finding and reconstruction on individual statistically significant hits. We therefore have to consider paradigms for statistical track finding that searches for tracks directly in low-level data.

The natural space to conduct template matching is the space of all tracks through the sensor array. This can be parameterized using a sphere bounding the sensor array. Every track passing through the sensor array is then represented by a pair of spherical coordinates, an entry time, and an exit time or a velocity, resulting in a 6-dimensional parameter space. 

Given this parameterisation, we can search for tracks by using template matching. Each track defines the position of a dark matter particle as a function of time, $\boldsymbol{r}(t)$; for a discrete measurement, we instead get a set of discrete positions, $\boldsymbol{r}_i$. For the $i^\mathrm{th}$ data sample and the $j^\mathrm{th}$ sensor, the signal template is
\begin{equation}\label{eq:template}
    \boldsymbol{f}_{ij} = \frac{\boldsymbol{r_{ij}}}{r^3_{ij}}.
\end{equation}

\begin{figure*}[htbp]
    \centering
    \includegraphics[width=0.95\textwidth]{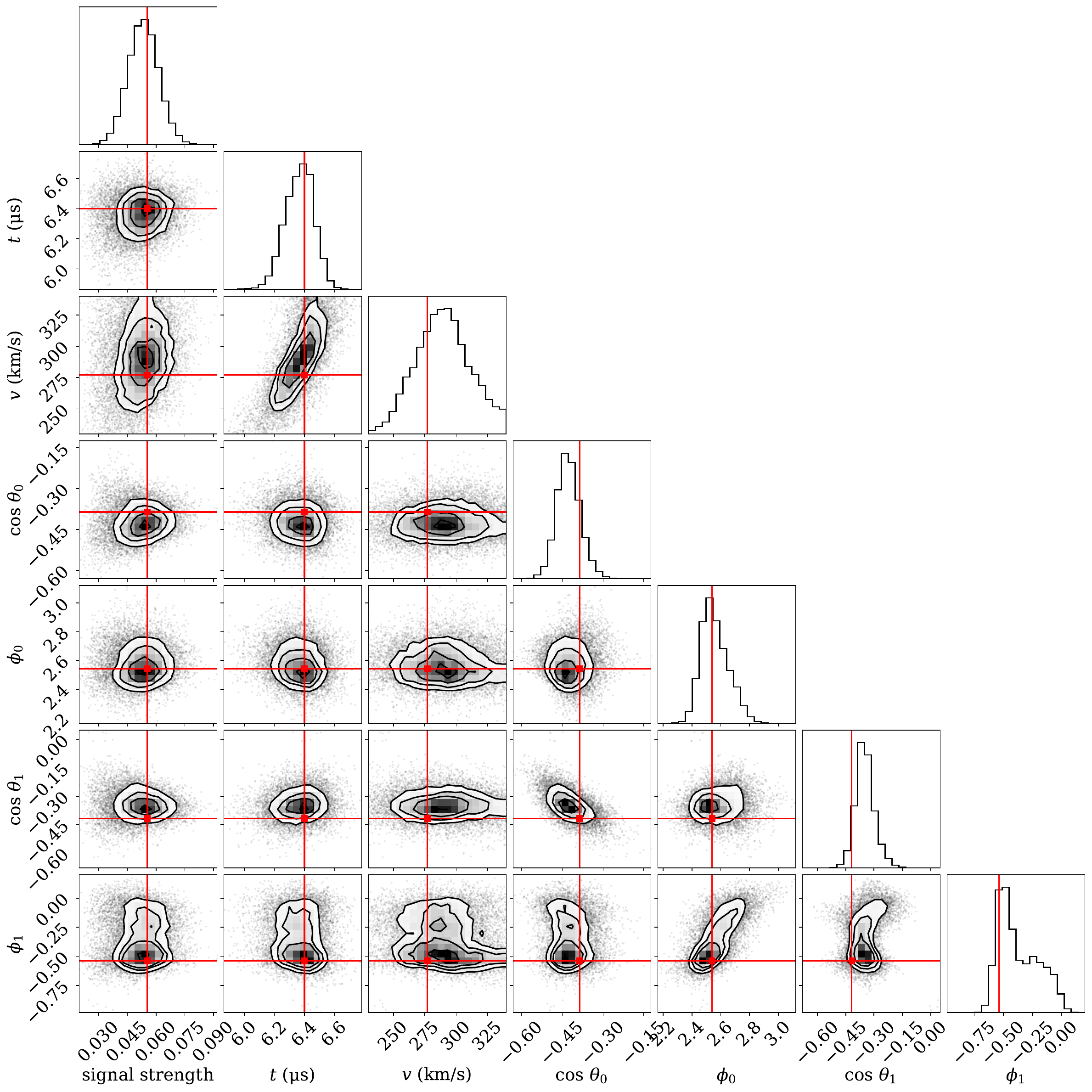}%
    \caption{A corner plot showing the posterior distribution of the temporal and geometric parameters and the signal strength produced using nested sampling, with the truth parameters indicated by the red crosses. The signal strength parameter is defined as~\cref{eq:signal_strength_prior}. $t$ and $v$ refer to the entry time and speed of the track, respectively. The geometric parameters are parameterised using the spherical coordinates of the entry and exit points of the track, as $(\cos \theta_0, \phi_0)$ and $(\cos \theta_1, \phi_1)$. It can be seen that the a posterior consistent with the truth value can be obtained in a full 6-dimensional search. The 6-dimensional 1, 2, 3 and 4\(\sigma\) contours are shown. In 2D histograms as shown here, these correspond to \(39\,\%,\: 86\,\%,\: 99\,\%\) and \(99.97\,\%\), because the probability content of \(\sigma\)-levels depend on dimensionality of the parameter space~\cite{Workman:2022ynf}. Illustration made with \texttt{corner} \cite{corner}.}\label{fig:bayes_corner_plot}
\end{figure*}

The signal strength is computed by convolving the template from each parameterised track with data. We can then compute the SNR by dividing this signal strength with the standard deviation of the noise. We can demonstrate this approach using simulated data from a $4^3$ sensor array, with white noise with a power spectral density of $6\times10^{-21}\,g/\mathrm{\sqrt{Hz}}$. Two-dimensional slices of track reconstruction with this simulated data are shown in~\cref{fig:2D_SNR}. We can see that this method can successfully reconstruct track parameters. 

One downside of this approach is that doing a likelihood scan in high dimensions is computationally expensive, as the number of likelihood computations is $N^d$ for a $d$-dimensional parameter space with $N$ bins per parameter. An alternative approach is to use nested sampling, where the scaling of computational cost with dimensionality depends on the information gain when the number of parameters is increased, but can be as low as $O(d^2)$ in some cases~\cite{Buchner2021NestedSM}.

When tackling this problem with nested sampling, analogously to~\cref{eq:template}, the likelihood function of the $i^\mathrm{th}$ data sample and the $j^\mathrm{th}$ sensor and $k^\mathrm{th}$ sensor axis is
\begin{equation}\label{eq:datapoint_likelihood}
    p_{ijk} \propto \exp\left({-\left(F_{ijk} - f_{ijk}(\boldsymbol{\theta})\right)^2/\left(2 \sigma^2_{ijk}\right)}\right),
\end{equation}
where $\boldsymbol{\theta}$ represents the dark matter track parameters being inferred, $F_{ijk}$ is the measured force on a sensor, and $f_{ijk}(\boldsymbol{\theta})$ is the expected force on a sensor given the parameters of the dark matter track~\cite{Qin:2023vlu}. The dark matter track parameters are $t$ and $v$, the entry time and speed of the track, $\cos \theta_0$ and $\phi_0$, the spherical coordinates representing the entry point of the track on the bounding sphere of the sensor array, and $\cos \theta_1$ and $\phi_1$, the spherical coordinates representing the exit point of the track on the bounding sphere of the sensor array. This assumes white noise where time samples are uncorrelated.

A flat prior can be chosen for all parameters except for interaction strength $\alpha$, which is not bounded. To alleviate this issue, we can define a signal strength parameter as
\begin{equation}\label{eq:signal_strength_prior}
    \zeta = \tanh \frac{\alpha}{\alpha_\mathrm{ref}},
\end{equation}
where $\alpha_\mathrm{ref}$ is a reference scaling parameter that can be adjusted. The signal strength, $\zeta$, is bounded within $(0,1)$, effectively serving as a prior on $\alpha$. $\alpha_\mathrm{ref}$ should be chosen such that $\alpha_\mathrm{ref} \gg \alpha_\mathrm{sens}$, the sensitivity of our experimental setup. This ensures that the prior is approximately flat near the values of $\alpha$ most relevant to discovering or setting limits on dark matter, while ensuring we have a proper prior.

Given \cref{eq:datapoint_likelihood}, the unnormalised log-likelihood of a time-series dataset would be
\begin{equation}
    \mathcal{L}(\boldsymbol{\theta}) = - \sum_{i, j, k} \frac{1}{2 \sigma^2_{ijk}}\left(F_{ijk} - f_{ijk}(\boldsymbol{\theta})\right)^2.
\end{equation}
The posterior distribution of a simulation of the same track used above in~\cref{fig:2D_SNR} is shown in~\cref{fig:bayes_corner_plot}. We can see that as with the template matching significance map shown in~\cref{fig:2D_SNR}, the true parameters are successfully recovered.

\subsubsection{The Look-elsewhere Effect and Detection Threshold}

One key issue with searches in any high dimensional parameter space is the look-elsewhere effect~\cite{Gross:2010qma}. In~\cite{Qin:2023urf}, we consider the look-elsewhere effect for searches of dark matter tracks through a $4^3$ sensor array using the framework of Gaussian random fields. This is done because the computational expense of simulating a high-dimensional search precludes the typical approach where one computes the $p$-value as a function of the significance using large sets of Monte Carlo simulations~\cite{Gross:2010qma}.

Instead, the covariance between different sets of track parameters is directly computed using the signal template. This allows for the significance map to be sampled as a multivariate Gaussian distribution. The $p$-value curve from these samples is then extended by fitting the expected analytic form of the excursion probability of a Gaussian random field. This procedure yields a trial factor of $\sim10^{22}$ for a $1\unit{yr}$ search, necessitating a significance threshold much higher than the typical $5\sigma$~\cite{Qin:2023urf}.

Another approach to estimating the trial factor would be to use the ratio of the posterior and prior densities~\cite{Bayer:2020pva}. If one assumes a mono-modal Gaussian posterior, the trial factor can then be computed as
\begin{equation}
\begin{split}
    \log{N_\mathrm{trials}} &= -\log{\frac{V_{\mathrm{posterior}, \boldsymbol{\theta}}}{V_{\mathrm{prior}, \boldsymbol{\theta}}}}\\
    &= -\log{B} + \log{\frac{V_{\mathrm{posterior}, s}}{V_{\mathrm{prior}, s}}} + \log{\hat{\mathcal{L}}},
\end{split}
\end{equation}
where $\frac{V_{\mathrm{posterior}, \boldsymbol{\theta}}}{V_{\mathrm{prior}, \boldsymbol{\theta}}}$ is the ratio of posterior to prior volume in track parameters, $\frac{V_{\mathrm{posterior}, s}}{V_{\mathrm{prior}, s}}$ is the ratio of posterior to prior volume in the signal strength parameter, $B$ is the Bayes factor, and $\hat{\mathcal{L}}$ is the maximum log-likelihood, as given in~\cite{Qin:2023vlu}. This is an approximation; we can can see from~\cref{fig:2D_SNR} and~\cref{fig:bayes_corner_plot} that the posterior is not entirely Gaussian.

Using this approximate procedure, we can obtain a trial factor of $\sim10^{20}$~\cite{Qin:2023vlu}. Due to these high trial factors, for the sensitivity plots in the rest of this work, we use a threshold of $10\sigma$. This is approximately the significance threshold needed for a global $p$-value of smaller than $0.1$ for a one year exposure. We note that this threshold is substantially harder to achieve for designs from previous studies, where 5$\sigma$ was used~\cite{Carney:2019pza}. In practice, fiducialization could result in a smaller trial factor, though this is not evaluated in detail in this work.

\section{Experimental Realization of Sensor Array}\label{sec:experimental_realisation}

There are multiple approaches to the experimental realisation of a sensor array aimed at detecting heavy dark matter~\cite{Windchime:2022whs, Janse:2024kcn}. In this work, we discuss two possible approaches, namely magnetic levitation~\cite{Hofer:2022chf, Latorre:2022vxo, Timberlake:2019swe} and solid-state accelerometers based on micro-electromechanical systems (MEMS)~\cite{ErrandoHerranz2020MEMSFP, Krause:2012iud}. Magnetic levitation systems have the benefit of being able to use relatively large test masses while maintaining a low amount of mechanical damping, while MEMS-based systems can scale up to a large sensor array with relative ease. In addition to these approaches, we consider a future milestone where a large array with low mechanical damping is considered.

These approaches are evaluated as they can scale up to macroscopic test masses, can achieve low noise levels due to low mechanical damping, and can be scaled up into sensor arrays.

\subsection{Magnetic Levitation}

Levitation of mesoscopic objects via the Meissner effect is one way to construct a sensitive force sensor, and this approach is an active area of research~\cite{Hofer:2022chf, Vinante2019UltralowMD}. Such sensors can be realised by either levitating a superconductor in a magnetic field, or by levitating a permanent magnet on a superconductor. This approach has intrinsically low thermal noise due to the cryogenic temperatures required~\cite{Hofer:2022chf, Vinante2019UltralowMD}. Magnetic levitation, also called maglev, has been used to construct excellent force and acceleration sensors~\cite{Fuchs:2023ajk}, and the first search for ultralight dark matter using this kind of technology has recently been performed~\cite{Amaral:2024rbj}. Additionally, integrated chip-based approaches offer a pathway to scaling up to sensor arrays~\cite{Latorre:2022vxo}. Finally, recent work has shown pathways to a very high degree of quantum noise reduction is available for microwave and rf-domain readout of levitated masses~\cite{Richman:2023mak}. This approach is thus a prime candidate for enabling the detection of small impulse signals from heavy dark matter, and has been proposed as the POLONAISE experiment to search for ultralight dark matter~\cite{Amaral:2024rbj}. We expect that it would be possible to search for heavy dark matter with the same experimental setup using different data acquisition and analysis pipelines.

\subsection{MEMS}

Achieving extremely high quality factors with macroscopic masses is more challenging with MEMS-based sensors, due in part to the difficulty of manufacturing heavy masses attached to tensioned springs. However, this chip-based approach leverages semiconductor manufacturing techniques, making cost-effectiveness and the ability to scale to large sensor counts as a key advantage~\cite{ErrandoHerranz2020MEMSFP}. In addition, as MEMS platforms can be compatible with both free-space and fibre-coupled optical readouts~\cite{ErrandoHerranz2020MEMSFP, Malayappan2021EnhancingFO}, squeezed light sources can be used to probe them~\cite{Pooser2019TruncatedNI}. Such squeezed light readouts have been used in practice, making them a more mature technology than quantum noise reduction techniques for magnetomechanical sensors such as those described in~\cite{Richman:2023mak}. As such, while the individual sensors might not outperform magnetically levitated sensors in terms of mechanical damping or sensor mass, larger sensors and the reduction of quantum noise using squeezed light readouts make the MEMS optomechanical platform competitive with magnetic levitation. This strategy has been proposed to enable large scale accelerometer arrays for the Windchime project~\cite{Windchime:2022whs}.





\section{Sensitivity Projections}\label{sec:sensitivity}

\subsection{Sensitivity Projection Method}\label{ssec:sensitivity_methodology}

Sensitivity projections are made using the following semi-analytical method~\cite{Qin:2023vlu}:
\begin{enumerate}
    \item A sensor array with the appropriate configuration is instantiated. For large sensor arrays, the sensor array is divided into identical cubic sections, and one such section of the array is instantiated. 
    \item We choose the number of tracks, $N_\mathrm{tracks}$, to simulate. $N_\mathrm{tracks}$ velocities, $v_i$ are sampled from the dark matter velocity distribution, according to the standard halo model with a local density of $\rho_\mathrm{\chi} = 0.3\unit{GeV/cm^3}$~\cite{Aalbers:2018mfc, aalbers_2019_3551727, Freese:2012xd}.
    \item Isotropically distributed tracks are generated by uniformly sampling the geometric space of tracks described in~\cref{sec:track-finding}. If the sensor array is divided into sectors, the point of entry and exit of each sector that the track intersects is computed.
    \item For a given track, the closest approach distance to each sensor, $b_i$, is computed. If the track intersects multiple sectors, this is repeated for each sector. 
    \item The SNR is computed for each instantiated sensor as a function of $v_i$ and $b_i$, and summed in quadrature.
    \item Steps 4 and 5 are repeated for each track, and the detection probability given a SNR threshold is then computed. In this work, we use a threshold of $\mathrm{SNR} = 10$, as suggested by~\cite{Qin:2023urf} and discussed in~\cref{sec:data-analysis}.
    \item The sensitivity of the sensor array is computed using the detection probability and the dark matter flux.
\end{enumerate}
This method is used instead of a full Monte-Carlo simulation of raw data, due to the computational difficulties described in~\ref{sec:track-finding}. We consider all sensors to be uni-directional, and sensor arrays are set up such that all sensors are pointing in the same direction.



We consider quantum and thermal noise as the dominant sources of measurement noise. 
We choose to focus on these sources of noise as they are inherent to the experiment and cannot be fully eliminated. We compute the measurement noise analytically, following~\cite{Qin:2023vlu}. For a free-particle, in the absence of quantum noise reduction techniques such as squeezing and back-action evasion, the uncertainty in the position of a test mass is given by
\begin{equation}
    \delta x^2 \geq \frac{\hbar \tau}{m},
\end{equation}
where $\tau$ is the measurement time and $m$ is the mass~\cite{Caves:1980rv}. When force measurements are made by subtracting subsequent positions, the force uncertainty is
\begin{equation}
    \delta F_{\mathrm{SQL}}^2 = \frac{4 m \hbar}{\tau^3},
\end{equation}
and the impulse uncertainty is 
\begin{equation}\label{eq:impulse_sql}
    \delta I_{\mathrm{SQL}}^2 = \frac{4 m \hbar}{\tau}.
\end{equation}
In a terrestial laboratory realisation, the test mass will be a bounded particle. In this case, this approximation only holds if $\omega < 1/\tau$.

In addition to quantum uncertainty, a test mass would be subject to thermal noise, given by $I^2_{\mathrm{thermal}} = \Gamma_\mathrm{thermal} \tau$, where $\Gamma_\mathrm{thermal}=4 m k T \gamma$, $T$ refers to the temperature, and $\gamma$ refers to the mechanical damping rate~\cite{Carney:2019pza, Clerk:2008tlb}. Parameterising the noise with a generic quantum noise reduction parameter, $\xi$, we can express the total measurement noise as thus:
\begin{equation}
    \delta I_{\textrm{noise}}^2 = \frac{4 m \hbar}{\tau\xi^2} + \Gamma_\mathrm{thermal} \tau.
\end{equation}
In this equation, we sum the squares of the two independent sources of noise, with quantum noise reduced proportionally by $\xi$, while leaving thermal noise untouched.

We can see that there exists a value of $\tau_\mathrm{opt} = {2}/{\xi}\sqrt{{\hbar m}/{\Gamma_\mathrm{thermal}}}$ that minimises the measurement noise. However, there is an additional consideration; in addition to minimising noise, the impulse signal should be largely contained within the measurement time $\tau$. Based on the integration bounds in~\cref{eq:momentum_transfer}, we enforce a constraint of $\tau \geq \frac{2b}{v}$. 
For large $\tau$, we instead use $\delta I_{\mathrm{SQL}}^2 = \hbar m \omega$, which is the expected noise model for a bound oscillator that is being measured for many periods~\cite{Clerk:2008tlb}. Ensuring this latter regime is piecewise continuous with the prior two regimes gives us the following approximation for total noise
\begin{equation}\label{eq:impulse_noise}
    \delta I_{\textrm{noise}}^2 =
    \begin{cases}
    \frac{2 m v \hbar}{b\xi^2} + \Gamma_\mathrm{thermal} \frac{2b}{v} &{\text{if }}\frac{2b}{v}\geq\tau_\mathrm{opt}\\
    \frac{4}{\xi}\sqrt{\hbar m \Gamma_\mathrm{thermal}}&{\text{if }}\frac{1}{\omega} > \tau_\mathrm{opt}>\frac{2b}{v}\\
    \frac{m \hbar \omega}{\xi^2} + \Gamma_\mathrm{thermal} \frac{1}{\omega} &{\text{if }}\frac{1}{4\omega}\leq\tau_\mathrm{opt}.
    \end{cases}
\end{equation}

The three conditions in~\cref{eq:impulse_noise} thus correspond to three different regimes where:
\begin{enumerate*}
    \item measurement noise can be minimized by choosing a smaller measurement time, but a measurement time of $2b/v$ is used to capture the full signal,
    \item measurement noise can be optimized using a measurement time longer than the minimum allowed measurement time of $2b/v$ but shorter than $1/\omega$, and
    \item the measurement time is long enough that the measurement noise is dominated by the ground state fluctuations of the oscillator.
\end{enumerate*}
It should be noted that this piece-wise expression is approximate around the transitions between regimes.

Finally, for MEMS-based setups, we also consider a noise model which accounts for laser power limits due to cooling constraints based on the methodology in~\cite{Lee:2025hkp}. This is detailed in~\cref{sec:power_limited_noise}.

\subsection{Sensitivity Projections}

\begin{table*}[htbp]
    \centering
    \caption{Parameters for different sensor array configurations.}\label{tab:parameters_windchime_arrays}
    \vspace{6pt}
    \begin{tabular}{|l||l|l|l|}
        \hline
        Parameter & Near-term MEMS & Near-term maglev & Future milestone \\
        \hline
        \hline
        
        \rule{0pt}{0.4cm}Mechanical quality factor \(Q_\mathrm{m}\) & \(10^7\) & \(10^7\) & \(10^{10}\) \\
        Resonance frequency \(\omega_\mathrm{m}\) & \(2\pi\times 20\unit{kHz}\) & \(2\pi\times 1\unit{Hz}\) & \(2\pi\times 20\unit{mHz}\) \\
        Sensor mass \(m_\mathrm{s}\) & \(20\unit{mg}\) & \(100\unit{mg}\) & \(100\unit{g}\) \\
        Sensor density & \(3.2\times10^3\unit{kg/m^3}\) & \(1.13\times10^4\unit{kg/m^3}\) & \(1.13\times10^4\unit{kg/m^3}\) \\
        Temperature \(T\) & \(15 \unit{mK}\) & \(15 \unit{mK}\) & \(15 \unit{mK}\) \\
        Quantum noise reduction \(\xi\) & \(10 \unit{dB}\) & \(0 \unit{dB}\) & \(15 \unit{dB}\) \\
        Sensor count & \(10\times10\times2\) & \(2\times2\times1\) & \(20\times20\times20\) \\
        Sensor array size & \(0.1\unit{m}\) & \(0.6\unit{m}\) & \(2\unit{m}\) \\
        Exposure & $1\unit{year}$ & $1\unit{year}$ & $5\unit{years}$ \\
        \hline
    \end{tabular}
\end{table*}

\begin{figure}[htbp]
    \centering
    \includegraphics[width=\columnwidth]{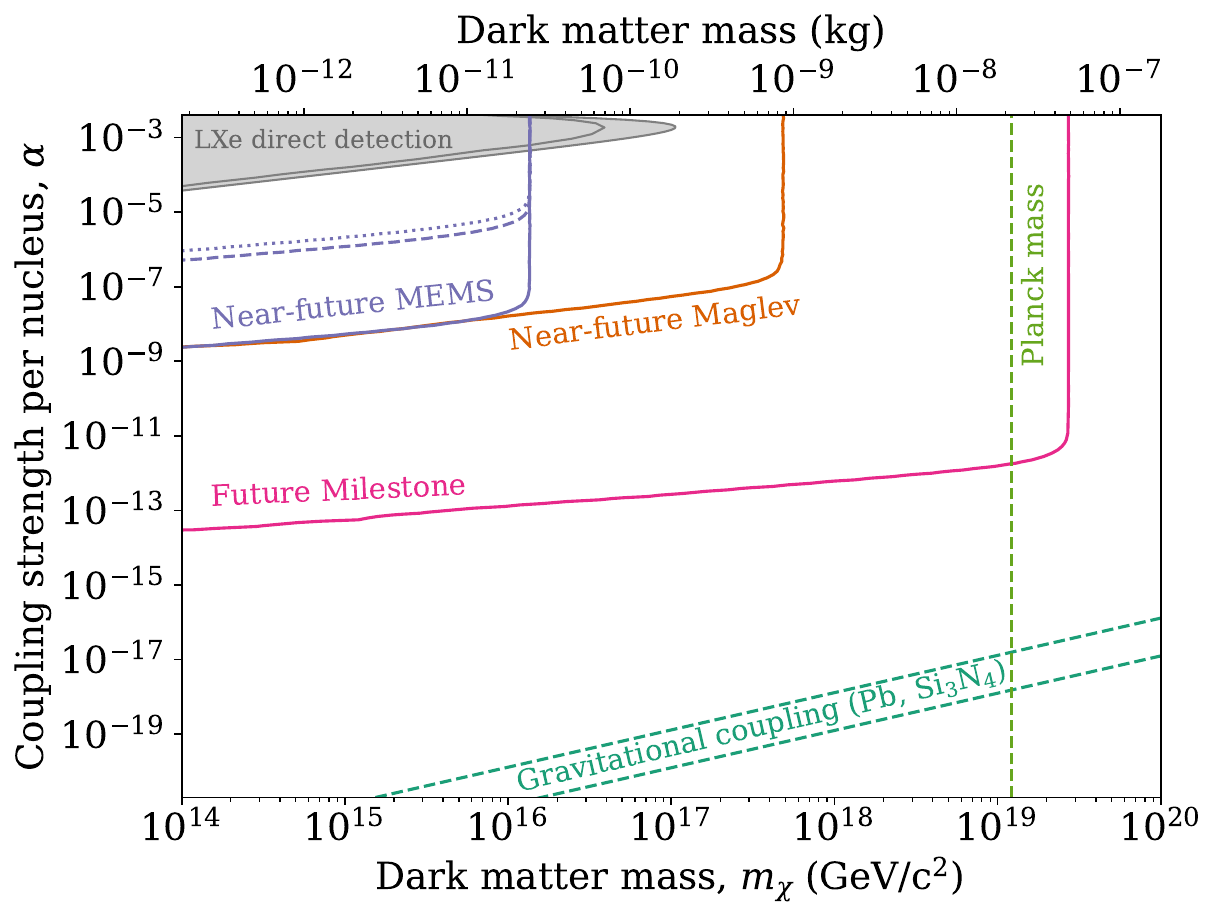}%
    \caption{Dark matter sensitivity of the different sensor configurations shown in~\cref{tab:parameters_windchime_arrays}. These sensitivity curves represent $90\,\%$ exclusion limits in the signal-free scenario, where there is no detectable dark matter for the given configurations. These sensitivity curves are shown alongside recast XENON1T and LZ limits~\cite{XENON:2023iku, LZ:2024psa}; details of the recasting procedure can be found in~\cref{sec:cross_section_alpha_conversion}. The value of $\alpha$ corresponding to the gravitational coupling is shown in the dark green dashed line. Two lines are shown due to material dependence of the equivalent gravitational coupling strength, this can be seen from~\cref{eq:force}, as the force is defined as a function of the number of nuclei instead of mass. The dotted and dashed lines for Near-future MEMS represent the detector sensitivity when the laser intensity is constrained by cooling power, with and without $10\unit{dB}$ of quantum noise reduction, respectively.}
    \label{fig:Windchime_sens}
\end{figure}

We consider three possible setups: a MEMS and a magnetic levitation experiment realisable in the near-term, and a third future milestone experiment. Finally, we also investigate what it might take to discover or rule out dark matter that only interacts gravitationally with the experimental setup. The parameters of the experimental setups are summarised in~\cref{tab:parameters_windchime_arrays}.

The near-term magnetic levitation temperature and mechanical quality factor $(Q_\mathrm{m})$ are based on~\cite{Hofer:2022chf}, while the mass is increased significantly to increase the sensitivity. The mechanical quality factor of the near-term MEMS setup is chosen to be $10^7$, as macroscopic mechanical oscillators with similar or higher quality factors have been demonstrated in mm-scale high-stress silicon nitride membranes~\cite{Chakram2013DissipationIU}; as such, we believe this is achievable for our application as long as tension is successfully retained in the silicon nitride spring. The temperature is chosen to be achievable with dilution refrigeration technology in large cm to metre scale volumes~\cite{CUORE:2014xzu, Hollister:2021lhg}. Limitations to the laser power due to heating is also considered for the near-term MEMS setup, based on a cooling power of $200\unit{\mu W}$, achievable with state of the art dilution refrigerators~\cite{Hollister:2021lhg} (see \cref{sec:power_limited_noise}). The near-term MEMS setup is given $10\unit{dB}$ of quantum noise reduction. 
This near-term quantum noise reduction for the optically accessible MEMS devices is based on the achieved 9 dB of squeezing demonstrated with rubidium vapor cell based free-space squeezed light sources~\cite{Dowran2018QuantumEnhancedPS, Glorieux2011QuantumCB}. In addition, the 15 dB of squeezing used in the future milestone is based on the record squeezing that has been achieved with any source~\cite{Vahlbruch:2016qwp}.
Finally, we take vibrational noise to be subdominant; this has been achieved at similar temperatures around $1\unit{Hz}$~\cite{vanHeck:2022evk}. The sensor material density is based on silicon nitride for the near-term MEMS setup as it is a favored material for high stress MEMS devices~\cite{Chakram2013DissipationIU}, and lead is used for the others as it is a dense type-I superconductor that has been demonstrated for chip-based magnetic levitation~\cite{Latorre:2022vxo}.

The sensitivity of the three setups, including the cases where the laser power constraints for MEMS are taken into account, can be seen in~\cref{fig:Windchime_sens}, compared with recast XENON1T and LZ limits~\cite{XENON:2023iku, LZ:2024psa}. Details of this the conversion between direct detection cross sections and the coupling strength, $\alpha$, can be found in~\cref{sec:cross_section_alpha_conversion}.

We can see that we can cover significant new parameter space with both proposed near-future setups even when limitations on readout laser power due to cooling power are considered, with the magnetic levitation setup being able to reach lower coupling strengths once these cooling power considerations are accounted for. We can also see that the future milestone setup can extend the reach in both coupling strength and dark matter mass. Reaching the gravitational coupling strength, however, remains a significant challenge even considering a setup with more futuristic parameters.

\section{Reaching the gravitational coupling strength}\label{sec:grav}

Using the same method shown in~\cref{ssec:sensitivity_methodology}, we can investigate what would be required to probe for Planck-mass dark matter gravitationally. We use the same parameters as the future milestone setup as detailed in~\cref{tab:parameters_windchime_arrays}, but scale up the array to be a $5\unit{m}$ cube of $100^3$ sensors, and increase the sensor mass to $600\unit{grams}$. Varying the amount of quantum noise while keeping all other parameters constant allows us to determine the amount of quantum noise reduction needed to set limits at the gravitational coupling strength.
\begin{figure}[htbp]
    \centering
    \includegraphics[width=\columnwidth]{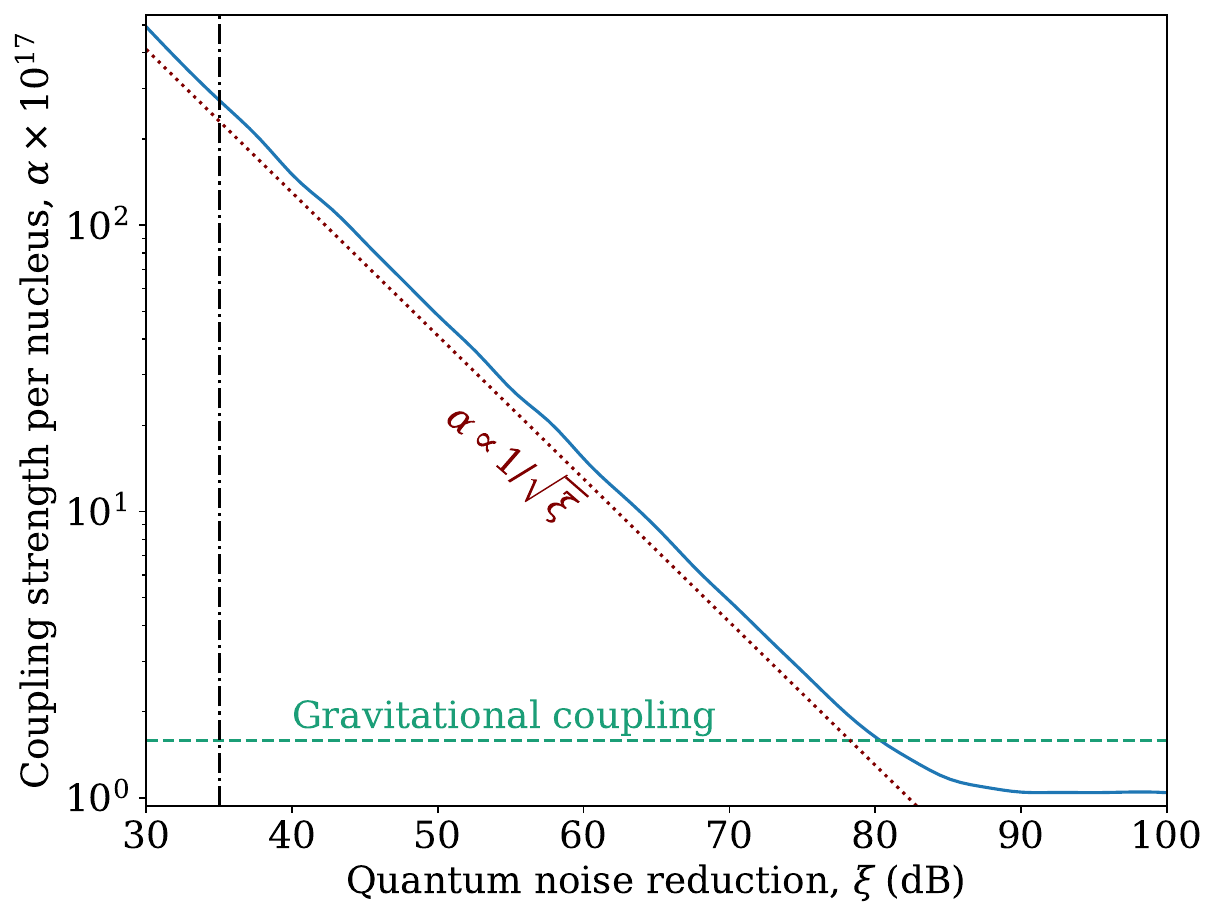}%
    \caption{The coupling strength $(\alpha)$ where the probability of a dark matter track above threshold exceeds $90\,\%$, as a function of the amount of quantum noise reduction $(\xi)$. The strength of the gravitational coupling for lead is indicated by the dark green dashed line. In addition, the expected scaling when quantum noise dominates is shown by the maroon dotted line. This corresponds to $\alpha\propto1/\sqrt{\xi}$, as can be seen from the second term in~\cref{eq:impulse_noise}. Above $\sim 90\unit{dB}$, we can see that the sensitivity is flat because thermal noise becomes dominant. The amount of quantum noise reduction needed to reach this thermally-limited regime from~\cite{Carney:2019pza} is shown by the vertical dot-dashed lines; the discrepancy is discussed in the text.}\label{fig:Windchime_alpha_scaling}
\end{figure}
The scaling of the sensitivity at the Planck mass as a function of the amount of quantum noise reduction is shown in~\cref{fig:Windchime_alpha_scaling}. We can see that approximately $80\unit{dB}$ of quantum noise reduction is required to reach the gravitational coupling, and $>85\unit{dB}$ is required to reach the thermal noise floor. This is in contrast to~\cite{Carney:2019pza}, where it is estimated that for sensors based on mechanical oscillators, $35\unit{dB}$ would be needed to reach the thermal noise floor where it becomes feasible to search for dark matter around the Planck mass using an array of impulse sensors. 

There are key differences in our estimation and prior work. Getting to the Planck mass regime in~\cite{Carney:2019pza} was examined for a 5$\sigma$ significance threshold and with a presumption of working near the SQL for resonance. Here we have found that 10$\sigma$ is crucial, combined with working with fixed frequencies and optimized bandwidths and with larger masses, consistent, for example, with findings of improved quantum noise reduction~\cite{Richman:2023mak}. While this could be a matter of increasing the size of the array, if instead we fix the array size, we find that the relevant parameters from prior work for detection push us out of the optimal sensing window as described by~\cref{eq:impulse_noise}.
In~\cite{Carney:2019pza}, quantum noise is computed as $\Delta I^2_{\mathrm{SQL}} = \hbar m \omega$, where $\omega$ is the resonance frequency of the mechanical oscillator. The issue is that while this is true if $\tau \gg \frac{1}{\omega}$, when the measurement time is close to or shorter than $\frac{1}{\omega}$, this approximation does not hold. As the characteristic time of our impulses is $\sim\frac{b}{v}$, where $b$ is the impact parameter and $v$ is the particle dark matter velocity, we expect measurement times to be $\lesssim\frac{1 \unit{m}}{200 \unit{km/s}}=5\times10^{-6} \unit{s}$ for a metre-scale experiment. We thus instead use the uncertainty given by~\cref{eq:impulse_sql}, where the characteristic frequency is $1/\tau$. As the resonance frequency is much lower than the optimal sampling frequency, this significantly increases the computed measurement noise arising from quantum uncertainty. We can see from~\cref{fig:Windchime_alpha_scaling} that the sensitivity indeed scales as expected from this regime.

\begin{figure}[htbp]
    \centering
    \includegraphics[width=\columnwidth]{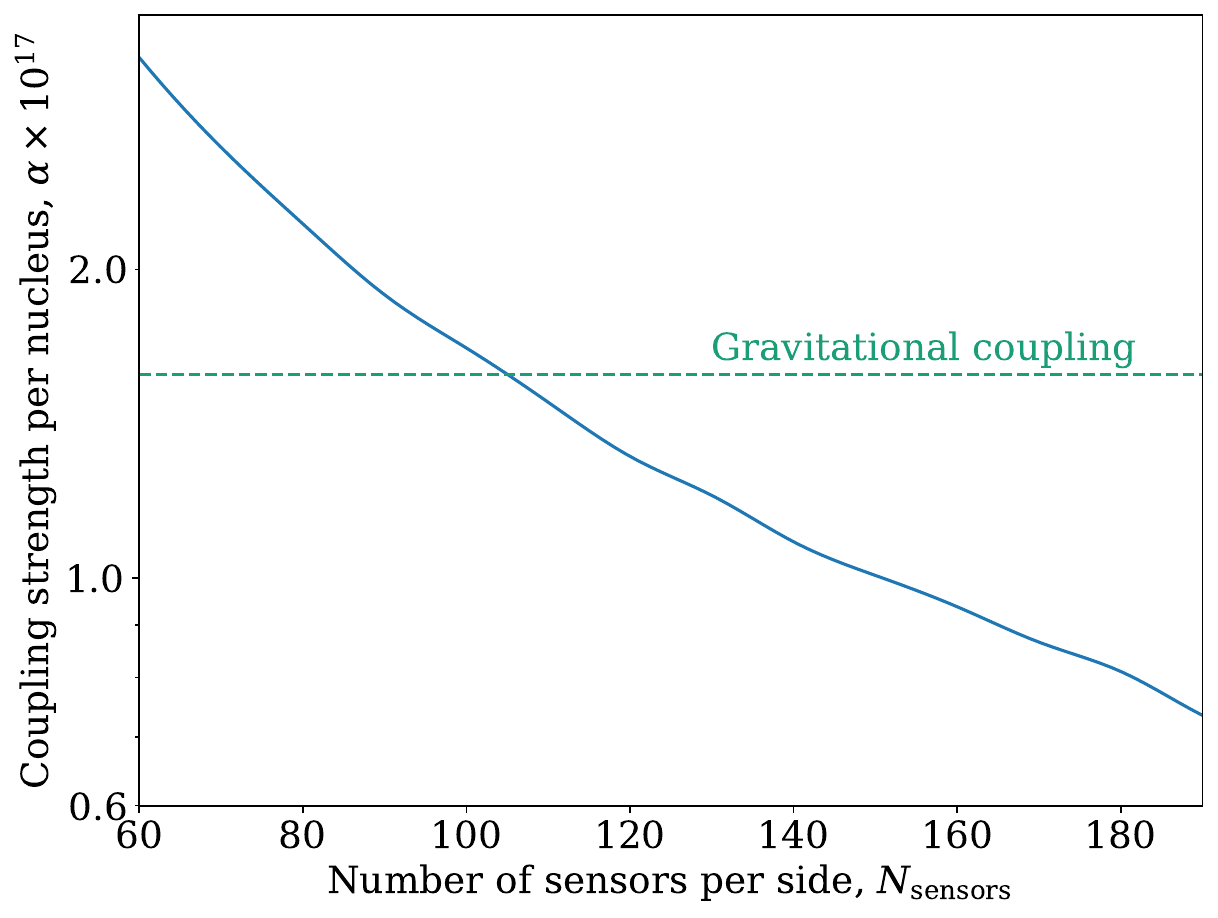}%
    \caption{The coupling strength $(\alpha)$ where the probability of a dark matter track above threshold exceeds $90\,\%$, as a function of the number of sensors per side. The strength of the gravitational coupling is indicated by the dark green dashed line.}\label{fig:Windchime_N_scaling}
\end{figure}

We are not aware of any path towards an experimental realisation of $\sim80\unit{dB}$ of quantum noise reduction; in addition, the setup involves a total levitated mass of $100^3\times600\unit{gram} = 600\unit{tonnes}$. As such, we view the gravitational coupling strength as a long-term goal that we should re-evaluate if new technologies change the feasibility of such an experimental setup, and not a concrete near-to-medium term goal.

We can also consider the scaling of the sensitivity with array sensor count, while keeping the other parameters including the spacing between adjacent sensors constant and using $80 \unit{dB}$ of quantum noise reduction. 
The scaling with array sensor count is shown in~\cref{fig:Windchime_N_scaling}. We can see that large increases in the size of a sensor array are needed to provide significant improvements in experimental sensitivity.

\section{Conclusion}\label{sec:conclusion}

In this work, we have explored the use of arrays of mechanical impulse sensors for the direct detection of ultraheavy dark matter that couples via a long-range force. We approached this problem from an experimentalist perspective, including consideration of the look-elsewhere effect, discussions of candidate technologies that can be used to probe this kind of dark matter, and using Monte-Carlo simulations to estimate sensitivities. We also developed statistical track-finding techniques that will be needed to conduct such a search of ultraheavy dark matter using arrays of mechanical impulse sensors.

In our sensitivity projections, we considered three experimental setups. The first two reflected experimental setups that are realisable with current or near-future technology, whereas the third represented a future setup that requires more research and development to be realised. The sensitivity of these setups was computed using a semi-analytical method that combined Monte Carlo sampling of dark matter tracks through the detector with analytic computation of the relevant signal-to-noise ratio.

We have shown that the setups we have described can cover significant new parameter space compared to existing experimental limits from XENON and LZ~\cite{XENON:2023iku, LZ:2024psa}. In particular, we expect sensor arrays of modest size sensing close to the quantum limit to be able to be two or more orders of magnitude better than existing limits if cooled to $\approx15 \unit{mK}$.

We found an even more pessimistic outlook for pure gravitational coupling searches than suggested in~\cite{Carney:2019pza}, demonstrating that even with very large sensor arrays of $~\sim 100^3$ sensors, it will take $\sim 80\unit{dB}$ of quantum noise reduction to approach the gravitational coupling limit, making it unreachable with current technology. While gravitational detection of Planck-mass dark matter is very difficult to achieve, experiments based on near-term realisable technology can cover large swathes of parameter space. Thus, we believe that searching for heavy dark matter using mechanical impulse sensors remains a promising avenue for future research.

\section*{Acknowledgements}

We would like to thank Charles Dyall, Sohitri Ghosh, Bas Hensen, Gerard Higgins, Zhen Liu, Andrew Long, and Tjerk Oosterkamp for technical assistance and discussions. This work is supported by the U.S. DOE Office of Science, High Energy Physics, QuantISED program (FWP ERKAP63) and the U.S. DOE Office of Science, Quantum Science Center.

\bibliography{main}

\clearpage
\appendix

\section{Conversion between cross section and coupling strength}\label{sec:cross_section_alpha_conversion}

We can recast direct detection limits into a long-range coupling strength parameter space by considering a long-range force that behaves analogously to the electromagnetic force. This procedure is based on the one outlined in~\cite{Qin:2023vlu}.

First, we note that the cross section between electrons and charged particles is given as~\cite{Harnik:2019zee}
\begin{equation}
\begin{split}
\sigma(E_r^{\min}, E_r^{\max}) &= \pi\alpha_\mathrm{EM}^2\varepsilon^2 \times \\
&\bigg[ \frac{m_e(E_r^{\max} - E_r^{\min})(2E_\chi^2 + E_r^{\max}E_r^{\min})}{E_r^{\max}E_r^{\min}(E_\chi^2 - m_\chi^2)m_e^2} \\
&- \frac{E_r^{\max}E_r^{\min}(m_\chi^2 + m_e(2E_\chi + m_e))}{E_r^{\max}E_r^{\min}(E_\chi^2 - m_\chi^2)m_e^2} \\
&\times \log\frac{E_r^{\max}}{E_r^{\min}} \bigg].
\end{split}
\end{equation}

We can consider a $B-L$ coupling for heavy dark matter, and analogously consider the cross section to be 
\begin{equation}
\begin{split}
\sigma(E_r^{\text{min}}, E_r^{\text{max}}) &= \pi\alpha_\mathrm{B-L}^2(A-Z)^2 \times \\
&\bigg[ \frac{m_n(E_r^{\text{max}} - E_r^{\text{min}})(2E_\chi^2 + E_r^{\text{max}}E_r^{\text{min}})}{E_r^{\text{max}}E_r^{\text{min}}(E_\chi^2 - m_\chi^2)m_n^2} \\
&- \frac{E_r^{\text{max}}E_r^{\text{min}}(m_\chi^2 + m_n(2E_\chi + m_n))}{E_r^{\text{max}}E_r^{\text{min}}(E_\chi^2 - m_\chi^2)m_n^2} \\
&\times \log\frac{E_r^{\text{max}}}{E_r^{\text{min}}} \bigg],
\end{split}
\end{equation}
where $\sigma$ is the per-nucleus cross section, $E_r^{\text{min}}$ is the minimum recoil energy, $E_r^{\text{max}}$ is the maximum recoil energy, $\alpha_\mathrm{B-L}$ is the $B-L$ coupling strength that is analogous to the fine-structure constant, $A$ is the atomic mass number, $Z$ is the atomic number, $m_n$ is the mass of the target nucleus, $E_\chi$ is the energy of the incoming dark matter particle, and $m_\chi$ is the mass of the incoming dark matter particle.

The inverse square force magnitude for the electromagnetic interaction is
\begin{equation}
    F_{\mathrm{EM}}=\frac{\alpha_{\mathrm{EM}} \hbar c q_1 q_2}{r^2 q_0^2}.
\end{equation}
The analogous equation for a $B-L$ coupling would thus be
\begin{equation}
    F_{\mathrm{B-L}} = \frac{\alpha_\mathrm{B-L} \hbar c (A-Z) q_\chi}{r^2}.
\end{equation}

To obtain an equation of the desired form in~\cref{eq:force}, we can define $\alpha=\alpha_\mathrm{B-L}(A-Z)$, and set the dark matter charge to unity. We can thus define $\beta$, a conversion factor between cross section and $\alpha$, as
\begin{equation}
\begin{split}
\beta(E_r^{\text{min}}, E_r^{\text{max}}) &= \pi \times \\
&\frac{m_n(E_r^{\text{max}} - E_r^{\text{min}})(2E_\chi^2 + E_r^{\text{max}}E_r^{\text{min}})}{E_r^{\text{max}}E_r^{\text{min}}(E_\chi^2 - m_\chi^2)m_n^2} \\
&- \pi\frac{E_r^{\text{max}}E_r^{\text{min}}(m_\chi^2 + m_n(2E_\chi + m_n))}{E_r^{\text{max}}E_r^{\text{min}}(E_\chi^2 - m_\chi^2)m_n^2} \\
&\times \log\frac{E_r^{\text{max}}}{E_r^{\text{min}}}.
\end{split}
\end{equation}
We then get
\begin{equation}
\begin{split}
    \sigma &= \beta (A-Z)^2 \alpha_\mathrm{B-L}^2\\
    &= \beta \alpha^2.
\end{split}
\end{equation}
We use $E_r^{\text{min}} = 6.4\unit{keV},E_r^{\text{max}}=43.5\unit{keV}$ for XENON1T~\cite{XENON:2018voc}, and $E_r^{\text{min}} = 5.5\unit{keV},E_r^{\text{max}}=54\unit{keV}$ for LZ~\cite{LZ:2022lsv}.

\section{Impulse sensitivity of power-limited optomechanical readout}\label{sec:power_limited_noise}

In this appendix, we show how the momentum sensitivity of a free-space MEMS depends on the incident laser power. We follow the analysis detailed in~\cite{Lee:2025hkp}, with the further inclusion of thermal noise, which contributes significantly in the regimes of interest here.

\begin{figure}
    \centering
\includegraphics[width=\linewidth]{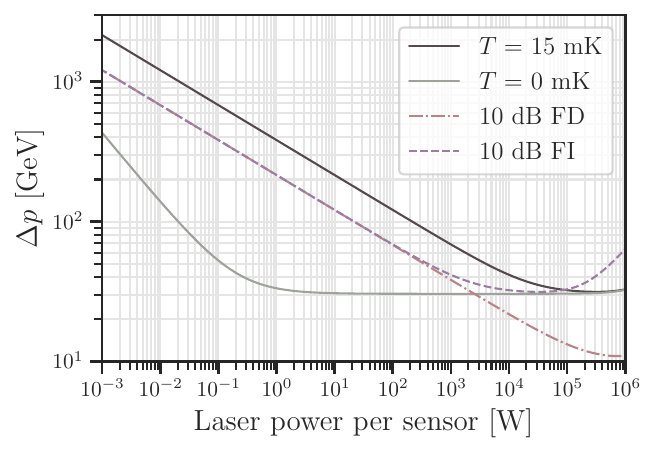}
    \caption{The momentum threshold as a function of the incident laser power, as given by Eq.~\eqref{eqn:pth}. We take the nominal MEMS parameters of Table \ref{tab:parameters_windchime_arrays}, and plot the threshold with no squeezing ($T = $ 15~mK), with 10 dB of frequency-dependent (FD) squeezing, as well as frequency-independent squeezing (FI) of the shot-noise $S^\mathrm{in}_{YY}$. For comparison, we also show the threshold in the absence of thermal noise, without squeezing ($T =$ 0~mK). }
    \label{fig:threshold_power}
\end{figure}

The MEMS acts as a free-space mirror whose position is inferred via homodyne measurement of the reflected laser light's phase. The mechanical response of the MEMS is described by the mechanical susceptibility 
\begin{align}
    \chi(\nu) = \big[m(\omega_\mathrm{m}^2 - \nu^2 - i \gamma \nu) \big]^{-1},
    \label{eqn:mech_susc}
\end{align}
which depends on the MEMS mass $m$, mechanical angular frequency $\omega_\mathrm{m}$, and damping rate $\gamma$. The latter two constants are related through the mechanical quality factor $Q_\mathrm{m} = \omega_\mathrm{m}/\gamma$.
The optomechanical coupling for such a system is given by $g = \sqrt{8
\omega_\mathrm{L} P_\mathrm{L}}$, where $\omega_\mathrm{L}$ and $P_\mathrm{L}$ are the laser's angular frequency and power, respectively~\cite{magrini2021real}. The force PSD at an angular frequency $\nu$ has contributions from both measurement and thermal noise, and is
\begin{equation}
\begin{split}
    S_{FF}(\nu) = &g^2 S_{XX}^\mathrm{in}(\nu) + \frac{1}{g^2 |\chi(\nu)|^{2}} S_{YY}^\mathrm{in}(\nu) \\
    &- 2 \frac{\mathrm{Re} \, \chi(\nu)}{|\chi(\nu)|^2} S_{XY}^\mathrm{in}(\nu) + 4 m \gamma T,
    \label{eqn:PSD}
\end{split}
\end{equation}
where the last term corresponds to thermal noise at a temperature $T \gg \nu$.
The first three terms in the PSD correspond to back-action $S_{XX}^\mathrm{in}$, shot-noise $S_{YY}^\mathrm{in}$,  and an interference term $S_{XY}^\mathrm{in}$, which depend on the squeezing parameters $r$ and $\theta$ through
\begin{equation}
\begin{split}
    S_{XX}^\mathrm{in}(\nu) &= \frac{1}{2} \left( \cosh 2r - \cos \theta(\nu) \sinh 2r \right), \\
     S_{YY}^\mathrm{in}(\nu) &= \frac{1}{2} \left( \cosh 2r + \cos \theta(\nu) \sinh 2r \right), \\
     S_{XY}^\mathrm{in}(\nu) &= -\frac{1}{2} \sin \theta(\nu) \sinh 2r.
     \label{eqn:Sii}
\end{split}
\end{equation}
The optimal momentum sensitivity is determined by the integral over angular frequencies of the inverse PSD
\begin{align}
    \Delta p = \left( \int \frac{d\nu}{2\pi S_{FF}(\nu)}\right)^{-1/2}.
    \label{eqn:pth}
\end{align}
Inserting Eqs. \eqref{eqn:mech_susc}, \eqref{eqn:Sii} and \eqref{eqn:Sii} into \eqref{eqn:pth}, we can calculate the momentum threshold as a function of the laser power, shown in Fig.~\ref{fig:threshold_power}. As in~\cite{Lee:2025hkp}, we allow for both frequency-independent squeezing $\theta(\nu) = \mathrm{const.}$, and frequency-dependent squeezing, where the squeezing angle $\theta(\nu)$ is varied in order to minimize the PSD at a given frequency. 

Given the cooling capacity of state of the art dilution refrigerators~\cite{Hollister:2021lhg}, we consider a total cooling power of 200~$\mu$W. With an array of 200 sensors, and assuming $0.1\%$ of the laser power is absorbed by the MEMS with highly-reflective coatings, the laser power per sensor must remain below 1 mW to avoid heating of the system, corresponding to a momentum sensitivity $\Delta p =$ 2.15~TeV in the absence of squeezing, and $\Delta p =$ 1.21~TeV for both frequency-independent and frequency-dependent squeezing, as shown in~\cref{fig:threshold_power}.

\end{document}

%% file: mydefs.tex
\newcommand{\unit}[1]{\, \mathrm{#1}}

%% file: authors.tex
\newcommand{\lbnl}{\affiliation{Physics Division, Lawrence Berkeley National Laboratory, Berkeley, California 94720-8153, USA}}
\newcommand{\maryland}{\affiliation{Joint Center for Quantum Information and Computer Science, University of Maryland, College Park, MD 20742}}
\newcommand{\nist}{\affiliation{Joint Quantum Institute, National Institute of Standards and Technology, College Park, MD 20742}}
\newcommand{\ornl}{\affiliation{Quantum Information Science Section, Computational Sciences and Engineering Division, Oak Ridge National Laboratory, One Bethel Valley Road, Oak Ridge, TN 37831, USA}}
\newcommand{\purduephys}{\affiliation{Department of Physics and Astronomy, Purdue University, West Lafayette, IN 47907, USA}}
\newcommand{\purdueoxide}{\affiliation{OxideMEMS Lab, Elmore Family School of Electrical and Computer Engineering, Purdue University, West Lafayette, IN 47907, USA}}
\newcommand{\qsc}{\affiliation{Quantum Science Center, Oak Ridge National Laboratory, Oak Ridge, TN 37831, USA}}
\newcommand{\ricecs}{\affiliation{Department of Computer Science, Rice University, Houston, TX 77005, USA}}
\newcommand{\ricephys}{\affiliation{Department of Physics and Astronomy, Rice University, Houston, TX 77005, USA}}

\author{Juehang~Qin\,\orcidlink{0000-0001-8228-8949}}\email{qinjuehang@rice.edu}\ricephys\qsc

\author{Dorian W.~P.~Amaral\,\orcidlink{0000-0002-1414-932X}}\ricephys
\author{Sunil~A.~Bhave\,\orcidlink{0000-0001-7193-2241}}\purdueoxide\qsc
\author{Erqian~Cai\,\orcidlink{0000-0003-0547-8727}}\ricephys
\author{Daniel~Carney\,\orcidlink{0000-0002-4269-8342}}\lbnl
\author{Rafael~F.~Lang\,\orcidlink{0000-0001-7594-2746}}\purduephys\qsc
\author{Shengchao~Li\,\orcidlink{0000-0003-0379-1111}}\altaffiliation[Now at ]{Department of Physics, School of Science, Westlake University, Hangzhou 310030, P.R. China} \purduephys
\author{Alberto~M.~Marino\,\orcidlink{0000-0001-5377-1122}}\ornl\qsc
\author{Giacomo~Marocco\,\orcidlink{0000-0001-7325-8190}}\lbnl
 \author{Claire~Marvinney\,\orcidlink{0000-0002-0289-8059}}
 \altaffiliation{This manuscript has been authored, in part, by UT-Battelle LLC, under contract DE-AC05-00OR22725 with the U.S. Department of Energy (DOE). The publisher acknowledges the U.S. government license to provide public access under the DOE Public Access Plan (\href{https://www.energy.gov/doe-public-access-plan}{https://www.energy.gov/doe-public-access-plan}).}
 \ornl\qsc
 \author{Jared~R.~Newton\,\orcidlink{0009-0001-2496-2613}}\purduephys\qsc
 \author{Jacob~M.~Taylor\,\orcidlink{0000-0003-0493-5594}}\nist\maryland
\author{Christopher~Tunnell\,\orcidlink{0000-0001-8158-7795}}\ricecs\ricephys